\documentclass[12pt, letterpaper]{article}
\usepackage[utf8]{inputenc}  

\usepackage[hidelinks,colorlinks=true,linkcolor=blue,citecolor=blue]{hyperref}
\usepackage[superscript,biblabel]{cite}
\usepackage[T1]{fontenc}
\usepackage{times}
\usepackage[top=1in,bottom=1in,left=1in,right=1in]{geometry} 

\usepackage{amsmath,amsthm,mathtools,amstext}
\usepackage{graphicx}
\usepackage{subfig}
\usepackage{float}
\usepackage{tikz}
\usepackage{caption}
\usepackage{authblk}

\begin{document}

\title{Theoretical investigation of slow gain recovery of quantum cascade lasers observed in pump-probe experiment}

\author[1]{Mrinmoy Kundu}
\author[1]{Aroni Ghosh}
\author[1]{Abdullah Jubair Bin Iqbal}
\author[1,*]{Muhammad Anisuzzaman Talukder}
\affil[1]{Department of Electrical and Electronic Engineering, Bangladesh University of Engineering and Technology, Dhaka-1205, Bangladesh}
\affil[*]{anis@eee.buet.ac.bd}
\date{}
\maketitle
\begin{abstract}
Time-resolved spectroscopy-based pump-probe experiments performed on quantum cascade lasers (QCLs) exhibit an initial fast gain recovery followed by a slow tail such that the equilibrium gain is not recovered in a cavity round-trip time. This ultra-slow gain recovery or non-recovered gain cannot be explained by only the intersubband carrier dynamics of QCLs. This work shows that the Fabry-Perot cavity dynamics and localized intersubband electron heating of QCLs are essential in ultra-slow and nonrecovered gain recovery. We developed a comprehensive model, coupling cavity dynamics to the intersubband electrons' thermal evolution. We employ a four-level coupled Maxwell-Bloch model that considers temperature-dependent scattering and transport mechanisms in calculating the gain recovery dynamics. If an intense pump pulse electrically pumped close to the threshold propagates in the forward direction after being coupled into the cavity, the reflected pump pulse will significantly deplete the gain medium while propagating in the backward direction. Additionally, we show that the intersubband electron sustains a localized high temperature even after the pump pulse has left, which affects the overall carrier dynamics and leads to an ultra-slow gain recovery process. At near-perfect reflectivity, we observe a gain depletion of 4\% for 2 mm QCL. We further demonstrate that an additional 10\% gain depletion of probe pulse is seen at a steady state when the laser is pumped at 1.6 times the threshold compared to the case where the hot electron effect is not considered. 
\end{abstract}

\section{Introduction}


The phenomenon of ultra-slow gain recovery in quantum cascade lasers (QCLs) can be better understood by using pump-probe studies, a common experimental technique\cite{liu2011femtosecond,choi2008gain,eickemeyer2002ultrafast}. Pump-probe experiments on lasers have been carried out for a long time to measure the gain dynamics \cite{sabbah2002femtosecond,mark1992subpicosecond,rohm2015dynamic}. In a pump-probe investigation, a very intense but narrow pulse is coupled to the laser cavity. This pulse is referred to as the pump pulse, and it significantly depletes the gain and does not allow significant gain recovery within its duration. A similar coherent and narrow but much less intense pulse, referred to as the probe pulse, is also injected into the laser cavity with a variable time delay to the pump pulse. As the intense pump pulse propagates, the population inversion changes because of various incoherent and coherent transport paths between quantized active and injector subbands, thus depleting the gain medium. The gain recovers exponentially at a characteristic time constant, often referred to as the gain recovery time\cite{Talukder2011}. The gain recovery time plays a vital role in many laser applications, such as the generation of short pulses by modelocking \cite{haus2000mode} and high modulation bandwidth for optical communication \cite{capasso2002quantum}.

Recent works have shown QCLs as a potential frequency comb (FC) generation device, especially in the terahertz frequency region \cite{Vitiell_2021,Li_2019}. Unfortunately, the QCL FC has several stability concerns that necessitate stabilizing techniques, including optical feedback, microwave injection locking, optical injection locking, and phase locking \cite{zhao_2020,consolino_nature_2019}. In particular, the short gain recovery time of the laser encountered in the QCL active region topology hinders stable pulse train creation and makes it challenging to perform classical pulsed passive modelocking \cite{wang_2015,tzenov_2018}, which is a particular sign of an incoherent multimode instability rather than a coherent frequency comb \cite{gordon2008multimode}. However, practical QCL FC devices yield broader linewidths due to several non-ideal circumstances like temperature drift, bias current variation, optical feedback, and other ambient noises. To date, theoretical explanations of the origin of QCL frequency comb instabilities are still lacking. Comprehensive gain dynamics and the hot electron effect studied in our work may provide insights into the direction of the development of such theories. 

Because of the ultrafast electron-longitudinal optical (LO) phonon interactions in QCL carrier transport\cite{weber1999intersubband}, the gain recovery of QCLs is very fast, on the order of a few picoseconds. The fast recovery of QCLs makes it challenging to achieve modelocking using conventional techniques \cite{gordon2008multimode,talukder2009analytical}. However, it allows QCLs to immediately follow changes in the injection current without relaxation oscillations, which is desirable for several applications, including high-speed free-space optical communications. Several pump-probe experiments have been performed to measure the gain recovery time and clearly understand the carrier dynamics of QCLs \cite{liu2011femtosecond,choi2008gain,eickemeyer2002ultrafast}. The time-resolved spectroscopy generally shows a speedy gain recovery on the scale of a few hundred femtoseconds at the beginning and then a relatively slower recovery on the order of picoseconds. The fast recovery is attributed to the rapid depopulation of the lower lasing level by emitting LO phonons and the coherent resonant tunneling to populate the upper lasing level. The slow recovery tail is attributed to the incoherent scattering transport through the quantum layers to extract carriers from the ground level of one period and populate the upper lasing level of the next period. Since the transport in QCLs depends on quantum mechanical designs, the details of the time-resolved gain recovery may vary from structure to structure.

Although the findings are broadly similar in the experimental and theoretical investigations of QCL gain recovery \cite{choi2008gain,eickemeyer2002ultrafast,Talukder2011}, there is a pronounced difference in the recovery tail obtained by Liu et al.~\cite{liu2011femtosecond}. The experimental findings show a long recovery tail, which reaches a steady-state value less than the equilibrium gain of the laser. The amount of non-recovered gain increases as the pump pulse energy increases. This unusually long gain recovery or the loss of gain is attributed to the slow electron transport through the injector region to populate the upper lasing level and the slow relaxation of electrons back to the active or injector region that is excited to continuum levels by the pump pulse. However, the observation of gain not being recovered on the order of a cavity round-trip time cannot be explained merely by the transport time through the quantum layers and the relaxation time of the electrons excited to continuum levels, as these transport times are often $\leq$ 1 ps \cite{choi2009time,jirauschek2014modeling}. A comprehensive model that includes electron excitation into the continuum levels, transport through the continuum levels, and the relaxation of electrons from the continuum levels to the confined levels has shown that the number of electrons excited to the continuum levels is too insignificant to be attributed to the long recovery tail or the non-recovered gain \cite{jamali2016comprehensive}. In Ref.~\citenum{jamali2016comprehensive}, Mahabadi et al.~concluded that one or more phenomena besides unipolar electron transport play an essential role in experimentally observed gain recovery.

This work shows that the Fabry-Perot cavity dynamics and the localized hot electron effect because of the high-intensity pump pulse help explain the ultra-slow gain recovery or the non-recovered gain in the pump-probe experiment. When electrical pumping to the gain medium is greater than the threshold, a significant portion of the reflected pump pulse could still deplete the gain medium, thus sustaining a non-recovered gain in the cavity round-trip time scale. Also, when an intense pump pulse propagates, the rapid transition of carriers between the active and injector regions can lead to a localized carrier heating effect. We develop a phenomenological electronic temperature model showing a 10--15\% rise in temperature, significantly affecting the carrier lifetimes and coherence time between the lasing levels. Coupling the temperature model with the Maxwell-Bloch equations and dynamically varying the carrier lifetimes, we find a much slower recovery time. The steady-state non-recovered gain increases with increasing current pumping ($p$) to the medium or pump pulse intensity ($E_p$). The fact that the gain difference is 10\% higher for the pumping parameter $p =$ 1.6 and 4\% higher for $E_p =$ 1.2 than it is when the temperature effect is ignored suggests that the thermal model plays a vital role in the dynamics of non-recovered gain. Our model, therefore, predicts slow gain recovery on the order of 35 ps or higher, which is consistent with the results of the earlier experiments.

In Sec.~\ref{Theoretical Model}, we present a four-level closed system approximation to model the injector, lasing, and depopulation dynamics of a QCL. We choose our reference for the QCL structure described in \citenum{ref26}. In Sec.~\ref{Hot electrons and electronic temperature}, we propose a phenomenological electronic temperature model that can account for the rapid localized electron heating effect and present calculation procedures of the scattering and coherence lifetimes. In Sec.~\ref{Results}, we analyze the results produced by our model and conclude.

\section{Theoretical Model} \label{Theoretical Model}
\begin{figure}[ht]
    \centering
    \includegraphics[width=0.8\textwidth]{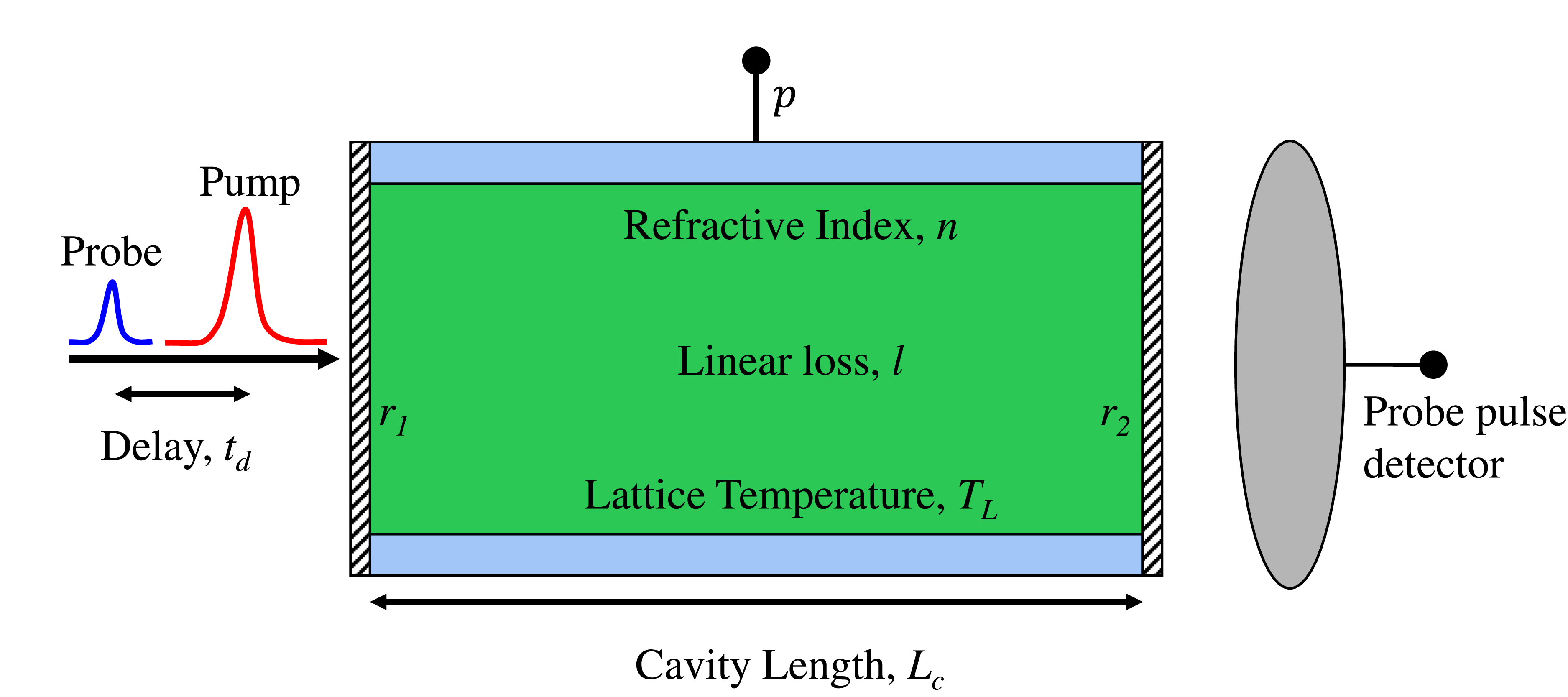}
    \caption{Schematic illustration of the simulated pump-probe experiment.}
    \label{fig:Schematic_pump-probe.}
\end{figure}
A schematic illustration of the theoretical model is shown in Fig.~\ref{fig:Schematic_pump-probe.}. The QCL cavity is electrically pumped by an external direct current source. As done in pump-probe experiments, we couple pump and probe pulses into the QCL cavity through one of the facets and the probe pulse is recorded from the other facet after a single pass. The length of the cavity is denoted by $L_c$. The lattice temperature and the refractive index of the cavity are given by $T_L$ and $n$, respectively. The reflectivities of the two cavity facets are $r_1$ and $r_2$. At the end of the cavity, a dectector is placed to record the probe pulse. The time delay between the probe and pump pulses is denoted as  $t_d$. Inside the cavity, the pulses experience attenuation due to various losses, all of which have been accounted for using a linear loss coefficient, $l$. The pump and probe pulses, coherent to the gain medium, are described by the following equations
\begin{subequations}
    \begin{align}
    E_{\rm pump} = E_p E ~{\rm sech}\left(\frac{t}{\tau}\right), \\
    E_{\rm probe} = \frac{E}{M}~{\rm sech}\left(\frac{t-t_d}{\tau}\right).
    \end{align}
\end{subequations}
Here, $E$ denotes the peak field of the pump pulse when $E_p =$ 1, with $E_p$ being a scaling factor used to vary the peak pump field, $E_{\rm pump}$, $M$ denotes the ratio of $E$ to the peak probe pulse, $\tau$ denotes the duration of the pulses (FWHM/1.763), and $t_d$ denotes the delay of the probe pulse relative to the pump pulse. Throughout this study, we assume $E =$ 3.66 $\times$ 10$^\mathrm{6}$ V/m, $\tau =$ 100 fs, and a fixed $M$ with a value of 30 for the probe pulse. With the exception of Sec.~\ref{Ep vary}, where we explore the impact of varying $E_p$ on gain recovery dynamics, we keep $E_p =$ 1 for all other cases. We calculate the intensity of the probe pulse at the output of the right facet after a single pass for a range of its delays from --2 ps to 35 ps considering the pump pulse. And unless otherwise stated, we have considered the QCL structure from Ref. \citenum{ref26} with applied electric field 60 \text{kV/cm} to calculate the carrier lifetimes throughout the paper.

\subsection{Closed four-level Maxwell-Bloch equations}
Previous studies have focused on describing the interaction between propagating pump and probe pulses and the QCL gain medium using either two-level coupled Maxwell-Bloch equations \cite{gkortsas2010dynamics} or multi-level extended density matrix formalism \cite{jamali2016comprehensive,Talukder2011}, but they did not consider cavity dynamics. To capture the dynamics of the injector and depopulation levels of a QCL active region more accurately, we have developed a four-level coupled Maxwell-Bloch model that comprehensively incorporates cavity dynamics. Additionally, we have included temperature-dependent carrier lifetimes in our model to account for the hot electron effect caused by the intense pump pulse. The coupled Maxwell-Bloch equations for a four-level closed system, appropriately normalized and utilized in our studies, are given by \cite{gkortsas2010dynamics,shimu2013suppression}

\begin{subequations}
\label{MB_eqs}
\begin{align}
&\frac{n}{c}\frac{\partial \Tilde{E}, \Tilde{e}_{\pm}}{\partial t}=\mp\frac{\partial \Tilde{E}, \Tilde{e}_{\pm}}{\partial z}-i\Tilde{\eta}_{E,e_\pm}-l\Tilde{E},\Tilde{e}_{\pm},\\
&\frac{\partial \Tilde{\eta}_{E, e_{\pm}}}{\partial t}=\frac{i}{2}[(\Tilde{\rho}_3-\Tilde{\rho}_2) \Tilde{E},\Tilde{e}_\pm + (\Tilde{\rho}^\pm_{32}-\Tilde{\rho}^\pm_{23}) \Tilde{E},\Tilde{e}_\mp]-\frac{\Tilde{\eta}_{E,e_\pm}}{T_2},\\
&\frac{\partial \Tilde{\rho}_3}{\partial t} = \frac{i}{2}(\Tilde{E}^{*}_{+}\Tilde{\eta}_{E_+} + \Tilde{e}^{*}_{+}\Tilde{\eta}_{e_+} + \Tilde{E}^{*}_{-}\Tilde{\eta}_{E_-} + \Tilde{e}^{*}_{-}\Tilde{\eta}_{e_-}-{\rm c.c.})-\frac{\Tilde{\rho}_3}{\tau_{3}}+\frac{\Tilde{\rho}_2}{\tau_{23}} + \frac{\Tilde{\rho}_1}{\tau_{13}} +\frac{\Tilde{\rho}_0}{\tau_{03}},\\
&\frac{\partial \Tilde{\rho}_2}{\partial t} = - \frac{i}{2}(\Tilde{E}^{*}_{+}\Tilde{\eta}_{E_+} + \Tilde{e}^{*}_{+}\Tilde{\eta}_{e_+} + \Tilde{E}^{*}_{-}\Tilde{\eta}_{E_-} + \Tilde{e}^{*}_{-}\Tilde{\eta}_{e_-} -{\rm c.c.})-\frac{\Tilde{\rho}_2}{\tau_{2}}+\frac{\Tilde{\rho}_3}{\tau_{32}} + \frac{\Tilde{\rho}_1}{\tau_{12}} +\frac{\Tilde{\rho}_0}{\tau_{02}},\\
&\frac{\partial \Tilde{\rho}_1}{\partial t} =-\frac{\Tilde{\rho}_1}{\tau_{1}}+\frac{\Tilde{\rho}_3}{\tau_{31}} + \frac{\Tilde{\rho}_2}{\tau_{21}} +\frac{\Tilde{\rho}_0}{\tau_{01}},\\
&\frac{\partial \Tilde{\rho}_0}{\partial t} =-\frac{\Tilde{\rho}_0}{\tau_{0}}+\frac{\Tilde{\rho}_3}{\tau_{30}} + \frac{\Tilde{\rho}_2}{\tau_{20}} +\frac{\Tilde{\rho}_1}{\tau_{10}},\\
&\frac{\partial \Tilde{\rho}_{32}^{\pm}}{\partial t} = \frac{i}{2}(\Tilde{E}^{*}_{\pm}\Tilde{\eta}_{E_\mp} + \Tilde{e}^{*}_{\pm}\Tilde{\eta}_{e_\mp} + \Tilde{E}^{*}_{\mp}\Tilde{\eta}_{E_\pm} + \Tilde{e}^{*}_{\mp}\Tilde{\eta}_{e_\pm})-\frac{\Tilde{\rho}^\pm_{32}}{\tau_{32}} + \frac{\Tilde{\rho}^\pm_{22}}{\tau_{23}},\\
&\frac{\partial \Tilde{\rho}_{22}^{\pm}}{\partial t} = -\frac{i}{2}(\Tilde{E}^{*}_{\pm}\Tilde{\eta}_{E_\mp} + \Tilde{e}^{*}_{\pm}\Tilde{\eta}_{e_\mp} + \Tilde{E}^{*}_{\mp}\Tilde{\eta}_{E_\pm} + \Tilde{e}^{*}_{\mp}\Tilde{\eta}_{e_\pm})-\frac{\Tilde{\rho}^\pm_{22}}{\tau_{23}} - \frac{\Tilde{\rho}^\pm_{22}}{\tau_{21}} + \frac{\Tilde{\rho}^\pm_{32}}{\tau_{32}},\\
&\tau_i^{-1} = \sum_{j\neq i} \tau_{ij}^{-1},\\
&\tau_{ij} =  \tau_{ij} (T_{\rm{ar}}),   T_2 = T_2 (T_{\rm{ar}}).
\end{align}
\end{subequations}
Here, $\Tilde{\eta_E}$ and $\Tilde{\eta_e}$ denote the normalized dielectric polarizations due to the pump and probe pulses, respectively, $\Tilde{E}$ and $\Tilde{e}$ denote the normalized envelopes of
electric fields of the pump and probe pulses, the quantities with a $+(-)$ subscript or superscript denote fields propagating in the positive (negative) $z$-direction, and the notation $\sim$ over symbols denotes that the corresponding parameter is normalized.

\begin{figure}[ht]
    \centering
    \includegraphics[width=0.5\textwidth]{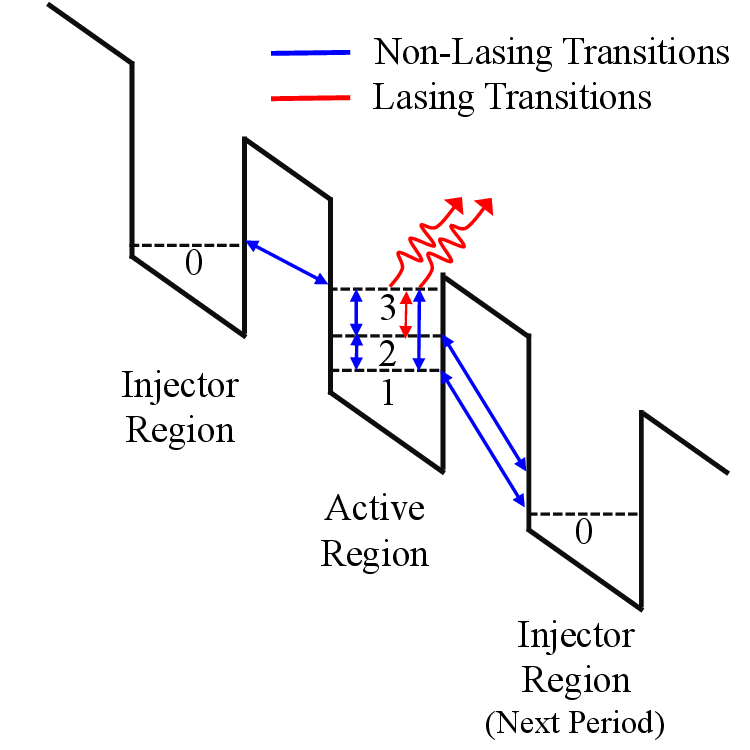}
    \caption{Schematic illustration of the carrier transport model for a four-level system. Blue arrows represent incoherent scattering mechanisms, and red arrows represent coherent carrier transport associated with photon emission. We use double arrows to signify that the carrier transport can be in either direction.}
    \label{fig:Schematic of the QCL 4 levels for a period}
\end{figure}

The normalization of the Maxwell-Bloch equations is discussed in Ref.~\citenum{temp_menyuk_2011}. The symbols $\Tilde{\rho_i}$ represent the diagonal density matrix elements for levels 0--3, as shown in Fig.~\ref{fig:Schematic of the QCL 4 levels for a period}. Furthermore, $\Tilde{\rho_{23}}$  and $\Tilde{\rho_{32}}$ correspond to the off-diagonal density matrix element related to the coherence between the lasing levels, and $\Tilde{\rho_{22}}$ represents the matrix element associated with the inversion grating. The parameters $\Tilde{E}$, $\Tilde{e}$, and $\Tilde{\rho}$ are assumed to vary slowly in both space ($z$) and time ($t$) according to the envelope function approximation. In addition, $c$ denotes the speed of light, $T_2$ denotes the coherence lifetime between lasing level 2 and 3, $\tau_i$ denotes the carrier lifetime of level $i$, and $\tau_{ij}^{-1}$ denotes the scattering rate from level $i$ to level $j$, where lifetimes depend on the electronic temperature of the active region $T_{\mathrm{ar}}$.
\subsection{Derivation of lasing threshold}
At steady state, we can assume that the forward and backward propagating electric fields and their interactions with the medium are the same. Therefore, we can write
\begin{subequations}
\begin{gather}
    \Tilde{E}_+ = \Tilde{E}_- = \Tilde{E},\\
    \Tilde\eta_{E, e_+} = \Tilde\eta_{E, e_-} = \Tilde\eta_{E, e},\\
    \Tilde\rho_{3,2,1}=\Tilde\rho_{3,2,1},\\
    \Tilde\rho^{+}_{32,22}=\Tilde\rho^{-}_{32,22}=\Tilde\rho_{32,22}.
\end{gather}
\end{subequations}
At steady state, using $\partial \Tilde{\eta}_{E, e_{\pm}}/{\partial t} = 0$ in
Eq.~(2b), the polarization becomes 
\begin{equation}
    \Tilde{\eta}_{E,e} = i\frac{T_2}{2}[(\Tilde{\rho}_3-\Tilde{\rho}_2) \Tilde{E},\Tilde{e} + (\Tilde{\rho}_{32}-\Tilde{\rho}_{23}) \Tilde{E},\Tilde{e}].
\end{equation}
If we substitute $\Tilde{\eta}_{E,e}$ into Eq.~(2a), we find the steady-state electric field evolution in the gain medium given by
\begin{equation}
   \left(\frac{\partial}{\partial z} + \frac{n}{c}\frac{\partial}{\partial t}\right)\Tilde{E},\Tilde{e} =\frac{T_2}{2}\left[(\Tilde{\rho}_3-\Tilde{\rho}_2) \Tilde{E},\Tilde{e} + (\Tilde{\rho}_{32}-\Tilde{\rho}_{23}) \Tilde{E},\Tilde{e}\right]. 
\end{equation}
Thus, the steady-state gain is given by $\Tilde{g} = T_2[(\Tilde{\rho}_3-\Tilde{\rho}_2)  + (\Tilde{\rho}_{32}-\Tilde{\rho}_{23})]/2$. Similarly, the steady-state density matrix elements can be determined as
\begin{gather}
\begin{bmatrix}
     1/\tau_{03} & 1/\tau_{13}  & 1/\tau_{23} & -1/\tau_{3}\\
     1/\tau_{02} & 1/\tau_{12}  & -1/\tau_{2} & 1/\tau_{32}\\
     1/\tau_{01} & -1/\tau_{1}  & 1/\tau_{21} & 1/\tau_{31}\\
     1 & 1  & 1 & 1\\ 
\end{bmatrix}
\begin{bmatrix}
     \Tilde\rho_{0} \\
     \Tilde\rho_{1} \\ 
     \Tilde\rho_{2} \\ 
     \Tilde\rho_{3} \\ 
\end{bmatrix}
=
\begin{bmatrix}
     0 \\
     0 \\ 
     0 \\ 
     \Tilde\rho_{0,\rm int}\\ 
\end{bmatrix},
\end{gather}
where the fourth row indicates the conservation of probability imposed on the density matrix elements and $\Tilde\rho_{0,\rm int}$ is the initial value of $\Tilde\rho_{0}$ when all carriers are in ground state. At the lasing threshold, we can set
\begin{subequations}
\begin{gather}
    \Tilde{E}, \Tilde{e}_+ = \Tilde{E}, \Tilde{e}_- = 0,\\
    \Tilde\rho^{+}_{32}=\Tilde\rho^{-}_{32}=0,\\
    \Tilde\rho^{+}_{22}=\Tilde\rho^{-}_{32}=0.
\end{gather}
\end{subequations}

With the matrix denoted as {\bf A}, the small signal gain $\Tilde{g_0}$ at the lasing threshold in the gain section can be expressed as $\Tilde{g_0} = \Tilde\rho_{0,\rm int} T_2(A^{-1}_{44} - A^{-1}_{34})/2$. Again, at the lasing threshold, the small-signal gain experienced by the propagating electric field has to overcome the sum of the linear loss in the cavity and reflection losses at the two facets. We can thus write
\begin{equation}
    \Tilde{g}L_c = \left[lL_c + \frac{1}{2}\ln\left(\frac{1}{r_1}\right)+\frac{1}{2}\ln\left(\frac{1}{r_2}\right)\right].
\end{equation}
Therefore, the threshold value of initial $\Tilde\rho_{0}$ to overcome the losses is
\begin{equation}
  \Tilde\rho_{0,\rm{int,th}} = \frac{2}{T_2(A^{-1}_{44} - A^{-1}_{34})} \frac{1}{L_c}\left[lL_c + \frac{1}{2}\ln\left(\frac{1}{r_1}\right)+\frac{1}{2}\ln\left(\frac{1}{r_2}\right)\right].
\end{equation}

\begin{figure}[htb]
    \centering
    \includegraphics[width=0.5\textwidth]{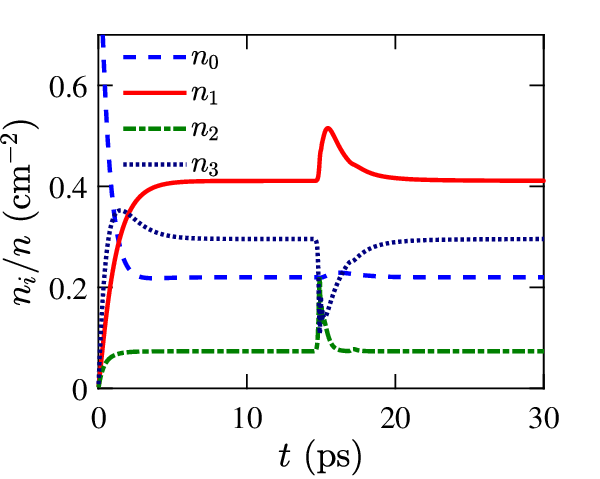}
    \caption{
    Time evolution of normalized carrier densities at the center of the cavity after injection of a pump pulse, where $n$ is the total carrier density and the system is at lasing threshold, i.e., $p = 1$.}
\label{fig: density}
\end{figure}

For the lasing system to operate at the threshold, the value of $\Tilde\rho_{0,\rm int}$ must be equal to or greater than $\Tilde\rho_{0,\rm int,th}$. By increasing the pumping rate, it is possible to surpass the lasing threshold. We introduce a parameter $p$ to quantify the strength of the current pumping applied to the medium, with $p =$ 1 indicating that the system is operating precisely at the threshold. Therefore, the expression for $\Tilde\rho_{0}$ can be defined as follows
\begin{equation}
    \Tilde\rho_{0,\rm int} = p\Tilde\rho_{0,\rm int,th}.
\end{equation}
With the initial condition of $\Tilde{\rho_0} = \rho_{0,\rm int,th}$, Fig.\ref{fig: density} shows that the normalized carrier densities reach steady state within only few picoseconds, and the lasing level become significantly depopulated as the pump pulse travels through the center of the cavity.
%

\section{Hot electrons and electronic temperature} \label{Hot electrons and electronic temperature}

When subjected to an intense pump pulse, depletion of QCL lasing levels and rapid depopulation of injector levels lead to increased incoherent carrier scatterings. These scatterings occur between bound states within the QCL and must conserve the total energy of the intrasubband electronic system. As a result, the electronic temperature becomes significantly higher than the lattice temperature $T_L$. This carrier heating phenomenon considerably impacts the QCL gain dynamics \cite{hot_electron_elsasser}. In addition, the carrier lifetimes and coherences vary with the instantaneous carrier temperature \cite{temp_menyuk_2011,lee1995}, a factor that was assumed to be constant in previous models \cite{jamali2016comprehensive,Talukder2011}.

\subsection{Modeling current density for incoherent carrier transport}
In addition to the coherent tunneling current, a substantial current density within the QCL medium arises from incoherent carrier transport. Since Eq.~(\ref{MB_eqs}) does not explicitly account for the coherence between energy levels other than the lasing levels, incoherent transport plays a significant role in the transport current and associated thermal phenomena. To estimate this current, we focus on a plane in the active region perpendicular to the bias current density that contains the depopulation level 1. In QCLs, intersubband transitions are primarily influenced by electron-LO phonon scattering, which significantly dominates over other scattering mechanisms. By considering the incoherent scattering paths from level 1 to the injector and the active region of the next period, we can express the incoherent current density as follows
%
\begin{figure}[htb]
    \centering
    \includegraphics[width=0.9\textwidth]{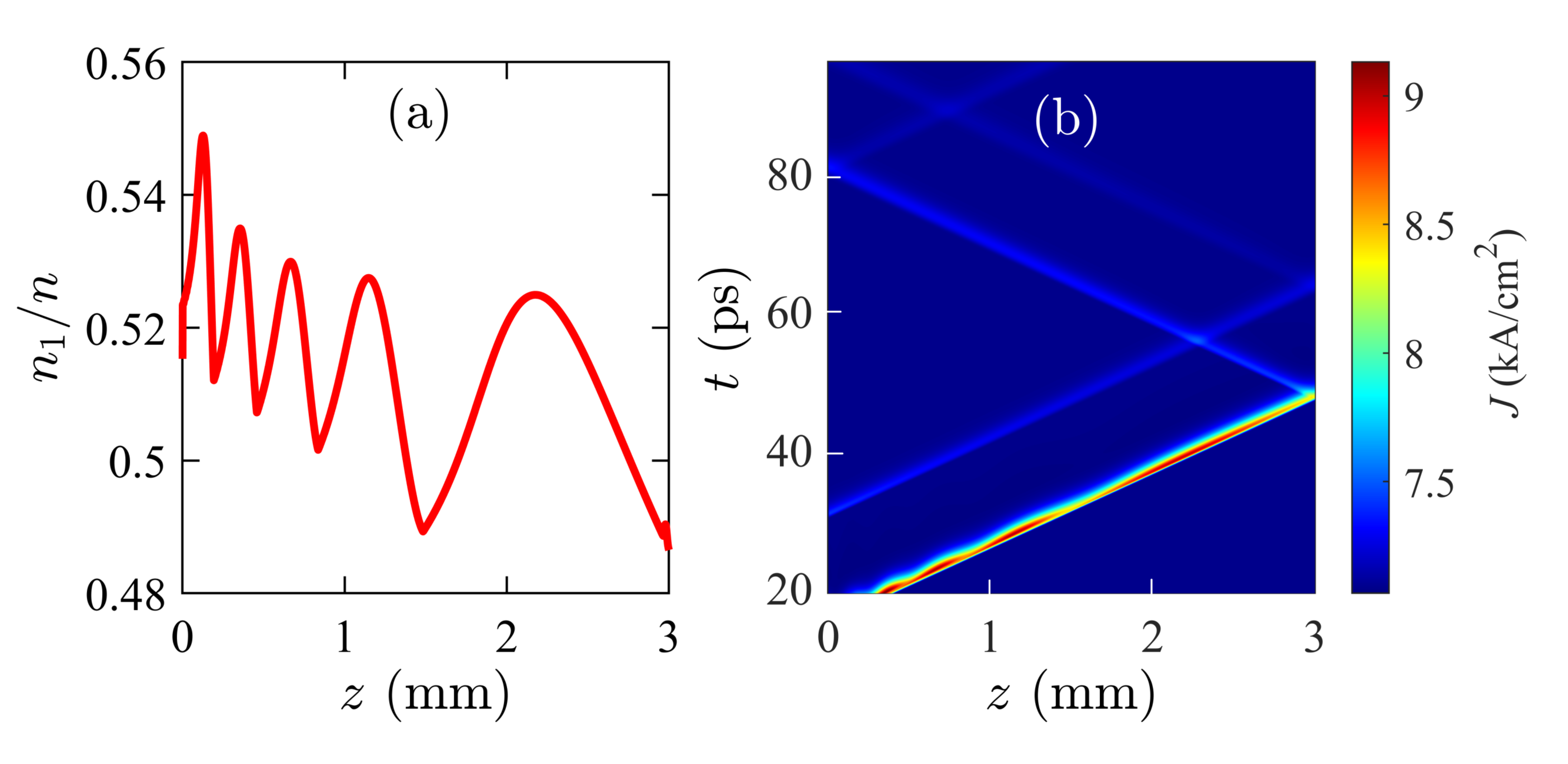}
    \caption{(a) Maximum normalized carrier density of level 1 with respect to the cavity position as the pump pulse propagates. (b) Spatial and temporal profile of incoherent current density $J$.}
    \label{fig: Jprofile}
\end{figure}
\begin{align}
\label{Incoherent_J}
J(n_1, T^e_{\rm ar})= q\left(\frac{n_1}{\tau_{10'}} + \frac{n_1}{\tau_{12'}} +\frac{n_1}{\tau_{13'}}\right).
\end{align}
Here, $\tau_{10'}$, $\tau_{12'}$, and $\tau_{13'}$ are the electron-LO phonon scattering times from the depopulation level 1 to the next period's injector and lasing levels. These carrier lifetimes will vary with temperature later in the model. Figure~\ref{fig: Jprofile}(a) shows the maximum value of the depopulation carrier density $n_1$ against the cavity length as the injected pump pulse propagates through the QCL medium. The maximum carrier density undulates due to the spatial hole burning, often prominent in QCLs \cite{spatialholeburning_gordon2008}. Although a single pulse is injected into the gain medium, the optical field is decomposed into two counter-propagating fields, creating a standing wave via interference and leading to a spatial pattern of excitation density and gain saturation \cite{spatialholeburning_gordon2008}. Figure~\ref{fig: Jprofile}(b) illustrates the spatial and temporal evolution of incoherent current density $J$, as defined in Eq.~(\ref{Incoherent_J}).
%
\subsection{Modeling evolution of electronic temperature}
Electronic energy distribution in each subband arises from the balance between the transition rates of several injection and energy relaxation channels via inter- and intra-subband electron--electron (e--e), electron--LO phonon, electron--impurity, and interface roughness scattering. The interplay between these processes can lead to different subband electronic temperatures $T^e_j$. For our case, we consider a one-temperature model where the thermal behavior of various QCL levels is effectively represented by one temperature in the active region $T^e_{\rm ar}$.

This approximation is justified since the e--e interactions with sub-picosecond time scale dominate over electron--phonon or electron--defect interactions. When an intense pump pulse excites the electrons in the quantum well subbands, nonequilibrium electronic distribution rapidly thermalizes to a Fermi distribution, resulting in a thermalized electronic distribution characterized by an effective temperature $T^e_{\rm{ar}}$ larger than the lattice temperature $T_L$. For device with high enough electron densities, e--e thermalization occurs in tens of femtoseconds in which no energy is transferred to the lattice so that $T^e_{\rm{ar}}$ remains greater than $T_L$ \cite{hot_electron_faist}. The hot electrons then lose energy to the lattice through intra- and inter-subband electron--LO phonon interactions.


We employ an energy balance equation by using the instantaneous and equilibrium current densities, $J$ and $J_{\rm eq}$, respectively, and the difference between the active region temperature $T_{\rm ar}$ and the lattice temperature $T_L$. We do not explicitly consider any lasing energy transfer between the electrons of the gain medium and the light field as they do not contribute to the thermal process.

The carriers in a subband can be considered as an electron gas confined in a 2D plane, implying that they have only two degrees of freedom. Hence, the carriers leave the active and injector region with an average thermal energy equal to $2\times\frac{1}{2}k_{B}T^e_j=k_{B}T^e_j$. The current density is the summation of transport through three scattering paths from the depopulation level 1 given by

\begin{subequations}
\begin{gather}
   J =\sum_i^3J_i= q\left(\frac{n_1}{\tau_{10'}} + \frac{n_1}{\tau_{12'}} +\frac{n_1}{\tau_{13'}}\right) =q\frac{n\tilde{\rho_1}}{K}\left(\frac{1}{\tau_{10'}}+\frac{1}{\tau_{12'}} + \frac{1}{\tau_{13'}}\right), \\\
   J_{\rm eq} = q\frac{n\tilde{\rho}_{1,\rm eq}}{K}\left(\frac{1}{\tau_{10}}+\frac{1}{\tau_{12'}} + \frac{1}{\tau_{13'}}\right). 
\end{gather}
\end{subequations}
Here, $K$ is the probability normalization factor. Assuming normalized density matrix elements $\tilde \rho_{i}$ being related to the corresponding actual matrix elements by a proportionality constant, $\tilde\rho_{i} = K\rho_{i}$, the conservation of probability implies $\sum n_i = \sum n\rho_i = \sum n \tilde\rho_i/K = \sum n \tilde\rho_{i,\rm eq}/K$, thus $K = \sum \tilde\rho_{i,\rm eq}$.

As electrons incoherently transport from level 1 to 0$'$, 2$'$ and 3$'$, they gain and lose energy through thermal relaxation with the lattice. Thus, the rate of change of excess energy in the plane of depopulation level 1 can be related as\cite{hot_electron_faist}
\begin{align}
    \frac{dE}{dt} = \frac{\sum_i (J_i-J_{i,\rm eq})\Delta_{i}}{q} - \frac{T_{\rm{ar}}-T_{L}}{\tau_{\rm eL}}nk_B,
\end{align}
where $\Delta_{i}$ is the change in electronic energy per carrier transport for the transition from 1 to 0$'$, 2$'$ and 3$'$,  $\tau_{\rm eL}$ is the thermal relaxation lifetime of electron-lattice interaction. $\Delta_{i}$ depends on the QCL structure and is calculated from 
the energy bandstructure based on a self-consistent Poisson-Schr\"{o}dinger model. When an intense pump pulse excites the electronic population, the Maxwell-Boltzmann distribution of the 2D electronic gas shifts toward the high kinetic energy region. These hot electrons with high $k$ values in the intra-subbands undergo two stages to cool down to lower $k$ values, firstly the cooling of the carriers due to emission of LO-phonons and consequently the decay of LO phonons, leading to an equilibrium with the lattice temperature $T_L$, with respective time constants $\tau_{\rm{eL-LO}}$ and $\tau_{\rm{LO}}$ \cite{hot_electron_elsasser}. As per the first order approximation for this two-stage process, we can lump the relaxation pathways into $\tau_{\rm eL} = \tau_{\rm LO}\tau_{\rm eL-LO} / (\tau_{\rm LO} || \tau_{\rm eL-LO})$, where $\tau_{\rm eL-LO} \sim$ 1 ps and $\tau_{\rm LO}$ is as high as 5 ps \cite{spagnolo2007hot_LOphonon_lifetime}. Generally, $\tau_{\rm eL-LO}$ bottlenecks the relaxation pathways and can significantly degrade $\tau_{\rm eL}$ \cite{tsai1993hot}. 
To explore a board range of variability, we take the value of $\tau_{\rm eL}$ on the order of 2--10 ps in the presence of an intense pump pulse.

In addition, QCL heterostructures possess strong anisotropy of thermal conductivity. The cross-plane conductivity component is much smaller than the in-plane one. This effect also plays a dominant role in overvaluing $\tau_{\rm eL}$. The presence of high-density abrupt sub-nanometer-sized interfaces causes phonon interference effects, which also inherently limit the heat extraction. \cite{hot_electron_faist}

Total thermal equilibrium energy per period can be given by $E = nk_BT_{\rm ar}$ so that
\begin{align}
    \frac{dT_{\rm ar}}{dt} = \frac{\sum_i (J_i-J_{i,\rm eq})\Delta_{i}}{qnk_{B}} - \frac{T_{\rm ar}-T_{L}}{\tau_{\rm eL}}.
\end{align}
Now, introducing normalized current density $\tilde{J_i} = J_iK/(qn)$, the time evolution of the active region temperature finally becomes

\begin{align}
    \frac{dT_{\rm ar}}{dt} = \frac{\sum_i(\tilde{J_i}-\tilde{J_{i,\rm eq}})\Delta_{i}}{K k_B} - \frac{T_{\rm ar}-T_{L}}{\tau_{\rm eL}},
\end{align}
with the initial conditions $T_{\rm ar}(t=0) = T_L$ and $\tilde J(t=0) = \tilde\rho_{1,\rm eq}(1/\tau_{10'} + 1/\tau_{12'} + 1/\tau_{13'}) $. For the initial conditions of the density matrix element in Eq.~(\ref{MB_eqs}), we set $\tilde\rho_{0}(t=0) = p\Tilde\rho_{0,\rm int,th}$, $\tilde\rho_{1}(t=0) = \tilde\rho_{2}(t=0) = \tilde\rho_{3}(t=0) = 0$, where $p$ is chosen suitably in the range of 1--2.
\begin{figure}[hbt]
    \centering
    \includegraphics[width=1.1\textwidth]{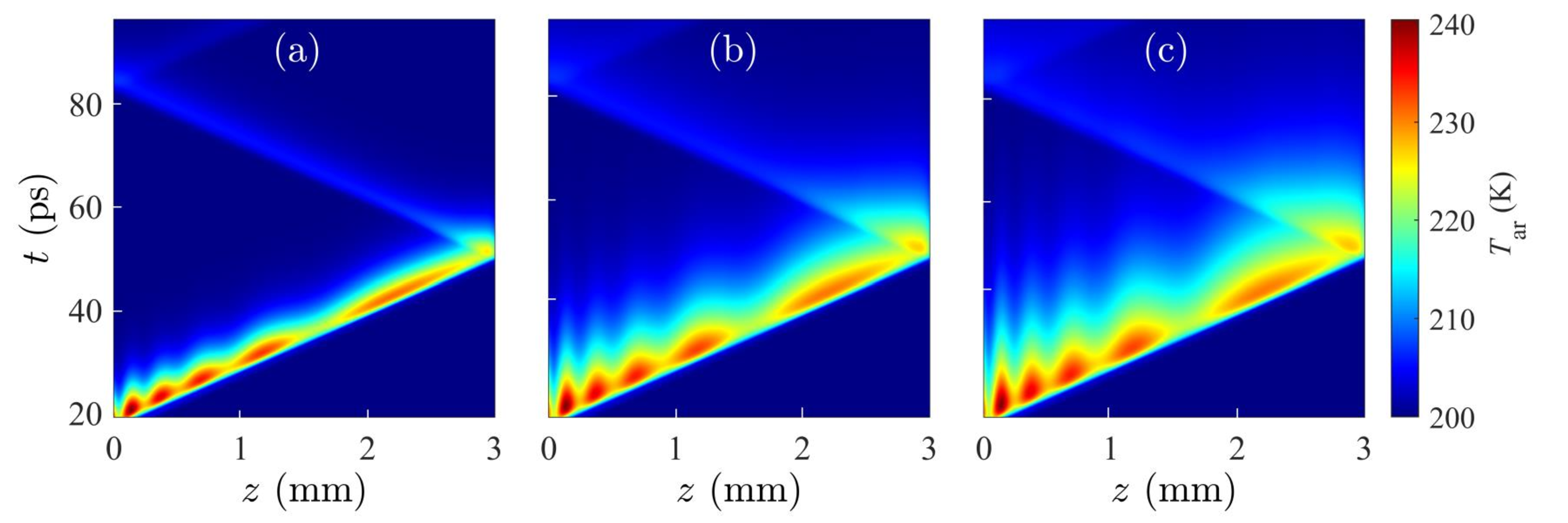}
    \caption{Spatial and temporal temperature profiles for (a) $\tau_{\rm eL} =$ 2 ps (b) $\tau_{\rm eL} =$ 6 ps, and (c) $\tau_{\rm eL} =$ 10 ps.}
    \label{fig:TempTile}
\end{figure}
%
\begin{figure}[t]
    \centering
    \includegraphics[width=1.1\textwidth]{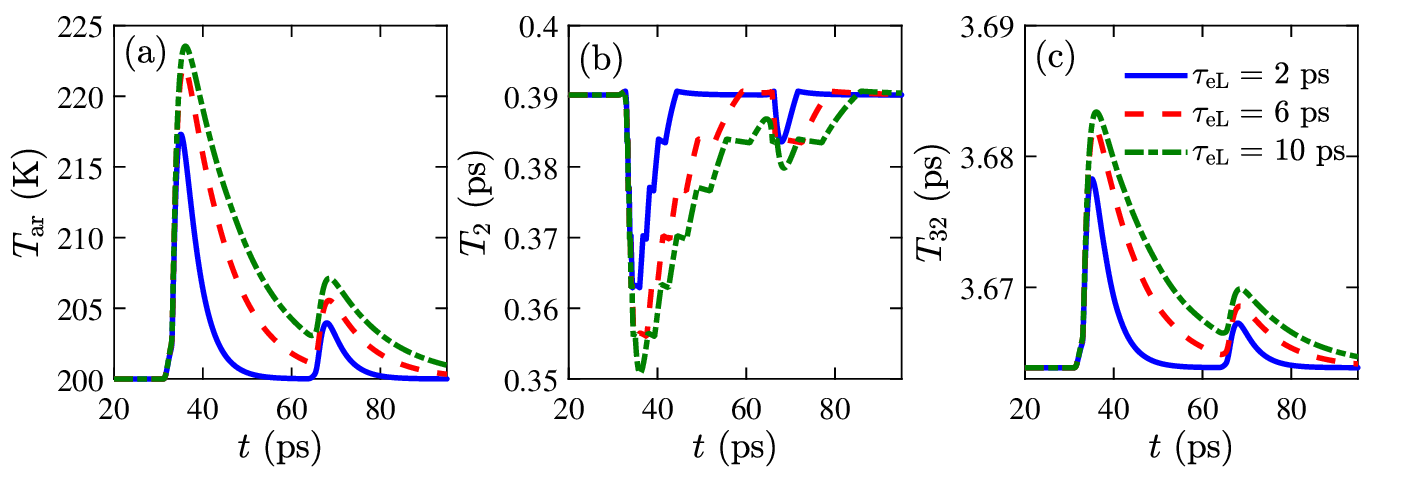}
    \caption{Dynamic variation of (a) intersubband temperature $T_{\rm ar}$ (b) coherence time $T_2$, and (c) scattering lifetime $T_{32} = (\tau_{32}^{-1} + \tau_{23}^{-1})^{-1}$ between lasing levels at the center of the cavity for different $\tau_{\rm eL}$.}
    \label{fig:TTile}
\end{figure}

Figure \ref{fig:TempTile} illustrates temperature profiles for three different thermal relaxation times when a probe pulse is injected into the QCL following an intense pump pulse. As $\tau_{\rm eL}$ increases, the temperature profile shows more smearing in the temporal domain, meaning that the localized heated regions sustain their electronic temperature for longer. The temperature at the center of the cavity also increases with $\tau_{\rm eL}$, as shown in Fig.~\ref{fig:TTile}(a). The probe pump injected into the gain medium later will encounter this localized heated region, leading to altered carrier dynamics with respect to the one with a constant homogeneous temperature profile model. Figure \ref{fig:TTile}(b) and (c) show the dynamic variation of coherence time $T_2$ and scattering lifetime $T_{32} = (\tau_{32}^{-1} + \tau_{23}^{-1})^{-1}$ between lasing levels. 
%
\subsection{Scattering lifetime calculation}
\label{Scat_time}
We have calculated scattering lifetimes for the QCL structure of Ref.~\citenum{ref26} designed on GaAs/AlGaAs material system. The applied electric field is 60 kV/cm, and $T_{L}$ is kept constant at 200 K. Intra-period and inter-period incoherent scattering lifetimes are calculated at different active region electronic temperatures considering e--e scattering and e--LO phonon scattering mechanisms where e--LO phonon scattering is usually more dominating \cite{ferreira1989,hartig1996}. Therefore, the intersubband scattering lifetime can be expressed as
\begin{equation} \tau_{xx'}=\ \left(\frac{1}{\tau^{\rm{e-e}}_{ xx'}}+\frac{1}{\tau^{\rm{e-ph}}_{xx'}}\right)^{-1}, \end{equation}
where $\tau^{\rm{e-e}}_{xx'}$ and $\tau^{\rm{e-ph}}_{xx'}$ are the carrier lifetimes for the transitions from $x$ to $x'$ due to electron-electron and electron-LO phonon scattering, respectively. 

The propagating electron wave packets lose phase coherence mainly due to intrasubband electron-LO phonon scattering, electron-electron scattering, and electron-interface roughness scattering\cite{Woerner_2004,wittman2008}. The coherence time between the coherently coupled energy levels 2 and 3 is calculated using
\begin{equation}\frac{1}{T_{2,23}}=\ \left(\frac{1}{T^{\rm e-e}_{2,23}}
+\frac{1}{T^{\rm e-ph}_{2,23}}+\frac{1}{T^{\rm e-ir}_{2,23}}\right). \end{equation}

\section{Results} \label{Results}
We have simulated a four-level QCL  schematically shown in Fig.~\ref{fig:Schematic of the QCL 4 levels for a period} to implement the gain recovery model. Using envelope function approximation of the Maxwell-Bloch equations, we have assumed that the pump and probe pulses are in resonance with the mid-IR spectral range of QCL emission wavelength. The key simulation parameters typical for QCLs \cite{gordon2008multimode, jamali2016comprehensive,choi2009time,modeling2014}, are listed in Table \ref{tab:my-table}. Electric fields of the injected pulses are allowed to propagate in both forward and backward directions and reflected from the edges by the characteristic reflection coefficients. As the propagating intense pump pulse depletes the gain medium, the rate of carrier injection into the upper lasing level 3 and extraction from the lower lasing level 2 will determine the gain recovery. The injection and extraction rates depend on scattering lifetimes and coherent tunneling rates between the injector and active region levels, which depend on the quantum mechanical design and the operating conditions, such as the applied electric field and temperature \cite{Talukder2011}.

\begin{table}[!ht]
\centering
\caption{Key Parameter Values}
\label{tab:my-table}
\begin{tabular}{lll}
\hline
\hline
\multicolumn{1}{c}{Parameter} & Symbol & Value \\ \hline
Cavity Length              & $L_c$    & 3 mm   \\
Facet Reflectivity            & $r_1$, $r_2$  & 0.53  \\ 
Refractive Index           & $n$      & 3.2   \\ 
Energy difference between levels           & $\Delta_{10'}$      & 42 meV   \\ 
& $\Delta_{12'}$      & 100 meV   \\ 
& $\Delta_{13'}$      & 230 meV   \\ 
Linear loss & $l$       & 1 cm\textsuperscript{-1}   \\
Emission Wavelength & $\lambda$       & 6.2 $\mu$m  \\
Lattice Temperature              & $T_L$    & 200 K   \\ 
Electric Field              & $F$    & 60kV/cm \\ \hline
\hline
\end{tabular}
\end{table}
\subsection{Effects of cavity parameters}
\begin{figure}[h!]
    \centering
    \includegraphics[width=0.5\textwidth]{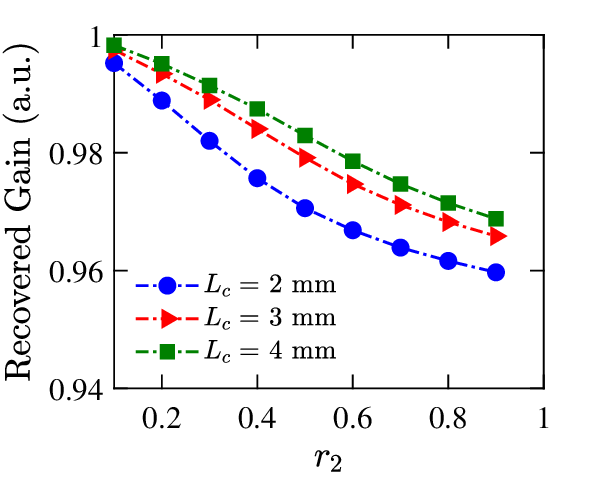}
    \caption{
    Recovered gain at \(t_d\) = 35 ps with varying  output facet reflectivity for different cavity lengths.}
    \label{fig: r2_gain}
\end{figure}
In this section, we explore the effects of parameters related to the cavity structure and QCL medium on the gain recovery dynamics. Firstly, in Fig.~\ref{fig: r2_gain}, we plot the recovered gain normalized by the equilibrium value with a variable reflectivity of the output facet for different cavity lengths. If the right facet reflectivity, $r_2$, is high, the reflected pump pulse will be sufficiently intense to deplete the gain medium while propagating in the backward direction. As a result,  the amount of recovered gain decreases. We also note that the recovered gain decreases for a smaller cavity as the reflectivity makes a more significant part of the total loss in a smaller cavity. When $r_2$ is small, e.g., 0.1, the recovered gain is almost unaffected by the cavity length since the intensity of the reflected pump pulse is equally small. However, when $r_2$ is large, $L_c =$ 2 mm yields the lowest recovered gain since a smaller cavity means a shorter round-trip time. Thus, an intense pump pulse will reflect back and forth through the medium in a smaller cavity, thereby depleting the gain more than it would do in a larger cavity.
%
\begin{figure}[htb]
    \centering
    \includegraphics[width=0.5\textwidth]{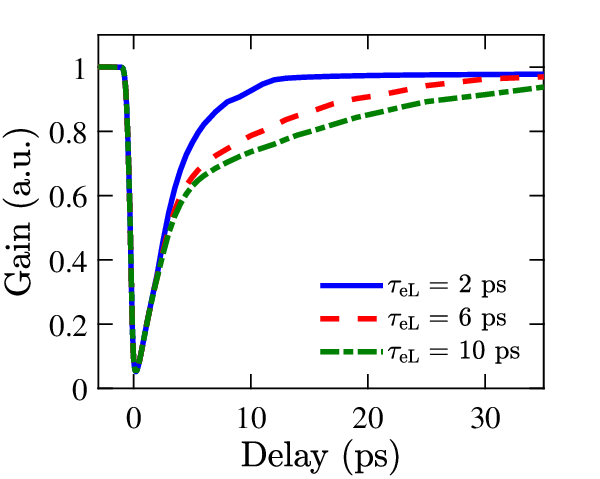}
    \caption{Normalized probe gain as a function of probe delay $t_d$ with different thermal relaxation time $\tau_{\rm eL}.$}
    \label{fig: tau_el}
\end{figure}

Furthermore, thermal relaxation parameter $\tau_{\rm eL}$ of the QCL medium directly correlates with the strength of the localized heating effect and the time the hot electronic temperature sustains. Hot electrons are cooled by the cascaded emission of optical phonons, resulting in a considerable amount of nonequilibrium optical-phonon population, which may induce major alterations in the system's carrier relaxation kinetics. Electron cooling rates in the active QCL region are drastically reduced in the presence of a considerable amount of hot-phonon feedback. \cite{hot_phonon}. Nonequilibrium phonons also result in enhanced electronic subband temperatures, as enhanced absorption of phonons effectively impedes electron energy relaxation \cite{nonequili_phonon}. As a result, for high values of $\tau_{\rm eL}$, the peak intensity of the probe pulses shows a more noticeable increase in the non-recovered gain along with a more slowly recovering tail shown in Fig.~\ref{fig: tau_el}. 
%
\subsection{Effects of pumping parameter}
%
\begin{figure}[h!]
     \centering
     \includegraphics[width=1\textwidth]{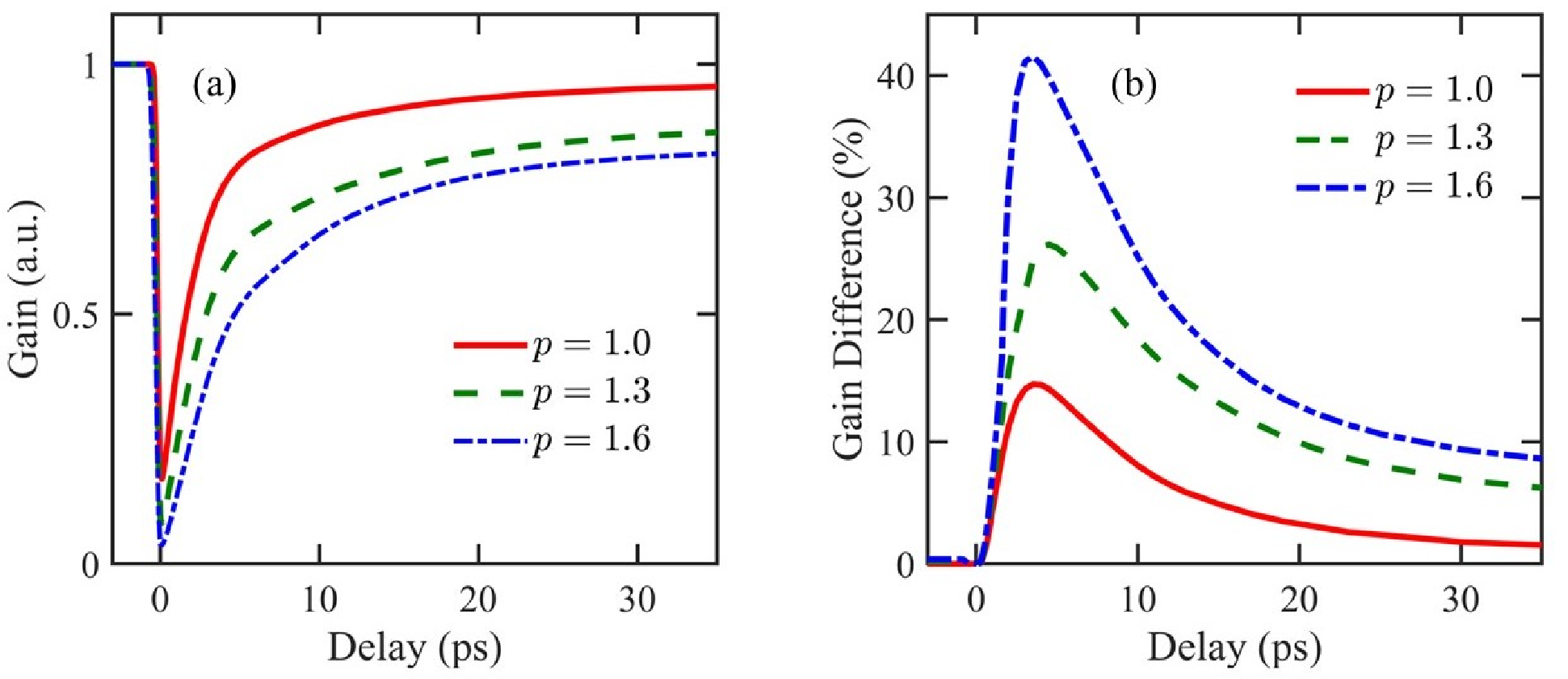}
    \caption{Effect of pumping parameter $p$ for (a) normalized probe gain vs probe delay $t_d$ (b) probe gain difference between the cases when hot electron effect is considered and not considered.}
     \label{fig: Effect of Pumping Parameter}
\end{figure}

Next, we inspect the effects of the current pumping parameter $p$. In Fig.~\ref{fig: Effect of Pumping Parameter}(a), we plot the peak intensity of the transmitted probe pulse through the right facet of the cavity after a single pass for different input currents pumping to the medium. In each case, the peak intensity of the probe pulses tends to reach a steady-state value less than the equilibrium. While the gain recovers exponentially with lasing levels carrier lifetime $\tau_{32}$, which depends on the intra-subband electronic temperature $T_{\rm ar}$ and the QCL electronic structure, the time-resolved spectroscopy also shows an ultra-slow recovery tail and a trend of reaching a steady-state less than the equilibrium value. Since the intensity of the reflected pump pulse increases with increased current pumping into the medium, the difference between the steady-state and equilibrium values increases too. We also note that the non-recovered gain does not increase linearly with $p$, as gain at $t_d =$ 35 ps decreases much drastically when $p$ is varied from 1 to 1.3. Thus, a slight increase in pumping current above the threshold can contribute significantly to the non-recovered gain dynamics.

To elucidate that the localized hot electron temperature affects the gain recovery dynamics, we also calculate the gain dynamics when the thermal model is turned off. We plot the difference in gain between these two cases in Fig.~\ref{fig: Effect of Pumping Parameter}(b). We note that the gain difference between the thermal model turned on and off is $\sim$ 10 \% when $p = $ 1.6, indicating a significant role of the thermal model on non-recovered gain dynamics.
%
\subsection{Effects of pump pulse parameter}
\label{Ep vary}

\begin{figure}[h!]
    \centering
    \includegraphics[width=1\textwidth]{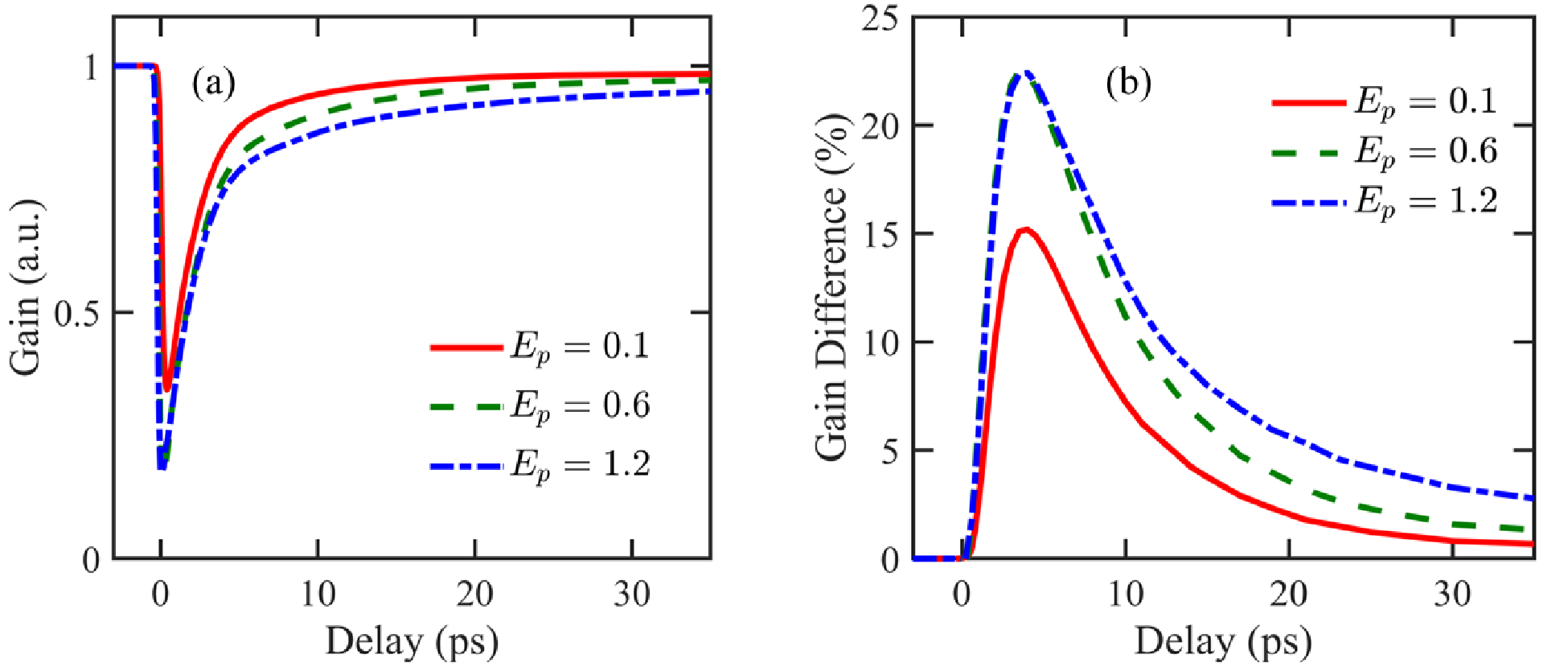}
    \caption{(a) Effect of pump field scaling parameter $E_p$ for (a) normalized probe gain vs probe delay $t_d$ (b) probe gain difference between the cases when hot electron effect is considered and not considered.}
    \label{fig:Ep_1}
\end{figure}
%
\begin{figure}[h!]
    \centering
    \includegraphics[width=0.5\textwidth]{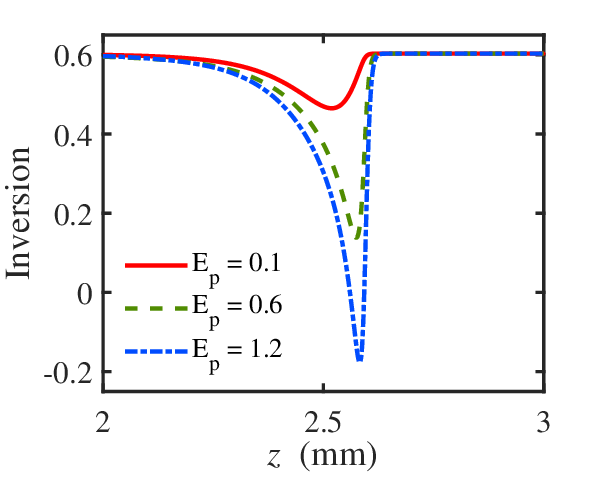}
    \caption{Population inversion  $[n_3(t) - n_2(t)]/[n_{3,\rm eq} + n _{2,\rm eq}]$ in the cavity with different pump intensity levels $E_p$.}
    \label{fig: inversion vs length}
\end{figure}

We plot the normalized probe gain for different values of the coupled pump pulse intensity $E_p$ in Fig.~\ref{fig:Ep_1}(a). The deviation of the steady-state gain from the equilibrium, i.e., non-recovered gain, increases as the pump intensity increases. The effect of the localized hot electron temperature is compared with the case when the thermal model is turned off in Fig.~\ref{fig:Ep_1}(b). It shows a maximum steady-state gain difference of 4\% when $E_p =$ 1.2, a similar indication that the thermal model explains a part of the non-recovered gain dynamics. \\
The increase of the non-recovered gain with the increase of the coupled pump intensity can be explained by the increase of the reflected and backward-propagating residual pump intensity, and hence, the increase of the gain depletion by the reflected pump pulse intensity. Figure ~\ref{fig: inversion vs length} shows the population inversion profile when the reflected pump pulse crosses the center of the cavity in the backward direction. The depleted population inversion due to the reflected pump pulse represents the loss of gain for the probe pulse.
%
\section{Conclusion}
In conclusion, the Fabry-Perot cavity dynamics and localized hot electrons play essential roles in the ultra-slow gain recovery in a pump-probe experiment of QCLs. An intense pump pulse significantly depletes the gain medium while traveling in the forward direction after being coupled and in the backward direction after being reflected from the edge. If gain depletion is significant, the probe pulse experiences a depleted gain medium, even if it is delayed by tens of picoseconds from the pump pulse. Under the influence of the intense pump pulse, incoherent scatterings between active and injector regions lead to a hot electron effect, affecting carrier transport through the quantized energy levels. As the intensity of the pump pulse or the output facet reflectivity increases, the amount of the non-recovered gain increases. The results qualitatively agree with experimental observations. In contrast to the conventional two-level model, the implemented four-level coupled Maxwell-Bloch system, including the cavity dynamics and hot electron effects, to model pump-probe experiments, help better understand the gain recovery profiles after the pump pulse depletes the QCL gain medium. 

\bibliographystyle{unsrt}
\bibliography{refs}

\end{document}